\documentstyle[epsfig]{texas}
\newcommand{\eqref}[1]{(\ref{#1})}
\newcommand{\dual}[1]{{}^*\!\left(#1\right)}
\newcommand{\coord}{co\-\"{o}r\-di\-nate}
\newcommand{\Coord}{Co\-\"{o}r\-di\-nate}
\newcommand{\coeff}{co\-\"{e}f\-fi\-cient}
\chardef\i="10
\begin{document}
\title{QUASI-STATIONARY BINARY INSPIRAL: PROJECT OVERVIEW}
\author{John T.~WHELAN}
\address{Institut f\"{u}r theoretische Physik, Universit\"{a}t Bern\\
  Sidlerstrasse 5, CH-3012 BERN, SWITZERLAND\\
  {\rm Email: whelan@itp.unibe.ch}}
\begin{abstract}
  I describe the current status of a collaboration with J.~D.~Romano,
  R.~H.~Price, and W.~Krivan to model the geometry of and
  gravitational radiation emitted by a binary system of compact
  objects in the regime where non-perturbative gravitational effects
  exist, but the rate of inspiral is still small relative to the
  orbital frequency.  The method of looking for a stationary spacetime
  which approximates the evolving solution is initially being tested
  on a simpler model with an additional translational symmetry.  This
  report consists of a general description of the method, followed by
  summaries of three techniques in varying stages of development: the
  simplification of the Einstein equations in the presence of two
  commuting Killing vectors which form a non-orthogonally-transitive
  symmetry group, the boundary conditions appropriate to the balance
  of ingoing and outgoing radiation needed to reconcile a stationary
  radiating solution with conservation of energy, and the treatment of
  gravitational waves far from the sources as linearized
  perturbations to the Levi-Civita spacetime.  The poster presentation
  with which this paper is associated is available on line at
  http://www-itp.unibe.ch/$\sim$whelan/poster.ps.gz and the current
  status of the project is described at
  http://www-itp.unibe.ch/$\sim$whelan/qsbi.html
\end{abstract}

\section{Introduction}

\subsection{Inspiral of Compact Object Binaries}

A pair of compact objects (black holes or neutron stars) in binary
orbit about one another is stable in Newtonian gravity.  In general
relativity, however, the system will emit gravitational radiation,
causing the bodies to spiral in towards one another.  The
gravitational radiation given off by this system is a prime candidate
for detection by  upcoming gravitational wave telescopes such as
VIRGO and LIGO.

When the compact objects are far apart, all gravitational effects are
weak and one can use the Post-Newtonian approximation, expanding in
powers of $GM/Rc^2$.  The final ``plunge'', where the objects cease to
orbit and collide rapidly, can be modelled by full general
relativistic three-plus-one supercomputer simulations.  An
intermediate late inspiral phase, where strong gravitational effects
are important, but the fraction of total energy lost to gravitational
radiation each orbit is small, cannot be handled by supercomputer
evolutions, which become unstable after several orbits.  To handle
strong gravity and slow inspiral, we need to use an approximation
scheme.  Modelling this intermediate phase will allow us to determine
the early gravitational wave signal as well as provide initial data
for the final plunge simulations.

\subsection{Quasi-Stationary Approximation}
  
The idea, initially proposed by Steven Detweiler \cite{det}, is that
if the inspiral is slow, the system is nearly periodic: after one
orbit, the objects have returned almost to their original locations,
and radiation which has moved out has been replaced with new radiation
of approximately the same shape.  If the objects' orbits are circular
rather than elliptical, the spacetime is nearly stationary.  If the
approximate orbital frequency is $\Omega$, moving forward in time by
$\delta t$ and rotating the resulting spatial slice by
$-\Omega\,\delta t$ will not change the picture very much.

Our approach is to replace the true spacetime, which has this
approximate continuous symmetry, with another solution to Einstein's
equations in which the symmetry is exact.  This ``stationary
quasi-solution'', which must somehow replace the energy lost in
radiation to prevent inspiral, should approximate the physical
``quasi-stationary solution'' over some time interval.  This replaces
a three-plus-one-dimensional numerical evolution with a
three-dimensional instantaneous solution, reducing greatly the
computing power required and hopefully escaping some of the numerical
instabilities associated with evolution.
  
To further simplify the problem, we will initially look for a
spacetime with an additional translational symmetry orthogonal to the
orbital plane.  This toy model of co-orbiting cosmic strings further
reduces the numerical problem to two dimensions while hopefully
retaining some of the qualitative features.

The rest of this paper consists of summaries of three techniques being
developed for this project, but which may prove useful in other
settings as well.  Section~\ref{sec:2KV} describes research done in
collaboration with Joseph D.~Romano \cite{2KV} on the refinement of a
formalism developed by Geroch \cite{geroch2} to simplify the Einstein
equations when the spacetime admits two commuting Killing vectors, as
is the case with the co-orbiting cosmic string model.
Section~\ref{sec:radbal} discusses the construction of a stationary
radiating solution to a wave equation without the use of external
forces by finding a preferred solution containing a balance of
incoming and outgoing radiation, the subject of a collaboration with
William Krivan and Richard H.~Price \cite{radbal}.
Section~\ref{sec:cyl} briefly touches on the concept of describing
small-amplitude gravitational radiation as a perturbation not to
Minkowski spacetime but to the Levi-Civita solution \cite{cs} for a
single cosmic string.

\section{QSBI I: Spacetimes with Two Killing Vectors \\
(Whelan and Romano \cite{2KV})}
\label{sec:2KV}

\subsection{Lack of a Block-Diagonal {\Coord} Basis}

The spacetime of two co-orbiting cosmic strings has two continuous
symmetries, which are described by two Killing vector fields.  One of
these, which we call $K_0$, represents the combination of time
translation and rotation about the orbital axis which leaves the
spacetime unchanged.  It can be thought of roughly in terms of
traditional cylindrical {\coord}s as $\partial_t+\Omega\partial_\phi$.
The other Killing vector simply corresponds to translation along the
strings and can be thought of as $K_1\sim\partial_z$.

In a numerical determination of the spacetime geometry, one seeks to
fix the {\coord} (i.e., gauge) information completely, and thus
calculate the minimum number of quantities necessary to define the
geometry.  It is desirable, of course, to choose a gauge which takes
advantage of the symmetries of the problem.  Following from the
example of static, axisymmetric spacetimes \cite{wald}, we might wish
to define two Killing {\coord}s $\{x^A\}$ ($x^0\sim t$, $x^1\sim z$),
supplemented by two other {\coord}s $\{x^i\}$ ($x^2\sim \rho$,
$x^3\sim\varphi\sim\phi-\Omega t$) on a subspace orthogonal to the two
Killing vectors, and bring the line element into a block-diagonal form
\begin{equation}
\lambda_{AB}(\{x^k\})dx^A dx^B+\gamma_{ij}(\{x^k\})dx^i dx^j.  
\end{equation}
For our set of Killing symmetries, however, this fails because the
symmetry group is not \emph{orthogonally transitive}, which means that
it is impossible to construct two-surfaces everywhere orthogonal to
both Killing vector trajectories.  The measure of this failure is a
pair of scalar fields $\{c_A\}$ given by
\begin{equation}
  c_A=\dual{K_0\wedge K_1\wedge\,d K_A}
  .
\end{equation}
The co-rotational Killing vector $K_0$ is not surface-forming
($K_0\wedge\,d K_0\ne 0$), which leads to the non-vanishing of $c_0$,
indicating a lack of orthogonal transitivity.
  
\subsection{Manifold of Killing Vector Orbits}

Although the metric components in any {\coord} basis will have
non-vanishing ``cross terms'' $\{g_{Ai}\}$ between the Killing and
non-Killing directions, it still possible to describe the geometry in
terms of two matrices $\{\lambda_{AB}\}$ and $\{\gamma_{ij}\}$ along
with the two scalars $c_0$ and $c_1$.  This done using a construction
due to Geroch \cite{geroch2} which defines a two-manifold
$\mathcal{S}$ of Killing vector orbits.  The {\coord}s $\{x^i|i=2,3\}$
are {\coord}s on this two-manifold, and any tensor on the original
four-dimensional spacetime manifold $\mathcal{M}$ which has vanishing
inner products with, and Lie derivatives along, the Killing vectors
corresponds to a tensor on the two-manifold $\mathcal{S}$.  In
particular, the tensors and scalars on $\mathcal{S}$ which describe
the spacetime geometry are
\begin{itemize}
\item The symmetric matrix of inner products $\lambda_{AB}=K_A\cdot K_B$
\item The metric on $\mathcal{S}$, which has components
  $\{\gamma_{ij}\}$ and corresponds to the projection tensor
  $g_{\mu\nu}-\lambda^{AB}K_{A\mu}K_{B\nu}$ on $\mathcal{M}$ (where
  $\{\lambda^{AB}\}$ is the inverse of $\{\lambda_{AB}\}$)
\item The two scalars $c_A$ which define the lack of orthogonal
  transitivity.
\end{itemize}

Given these objects, it is possible to define a non-{\coord} basis on
the four-dimensional spacetime by $e_A=K_A$ and
$e_{i}=\gamma_{ij}dx^j$.  In this basis the metric has components
\begin{equation}
  g_{AB}=\lambda_{AB}\,
, \qquad g_{Ai}=0\,
, \qquad g_{ij}=\gamma_{ij}\,
  ;
\end{equation}
the $\{c_A\}$ give the commutation {\coeff}s of the non-commuting
basis vectors:
\begin{equation}
  [e_2,e_3]\propto \lambda^{AB}c_A K_B\,.
\end{equation}

\subsection{Einstein Equations}

Geroch's original paper contained four partial differential equations
involving $\{\gamma_{ij}\}$, $\lambda_{00}$, $\lambda_{01}$,
$\lambda_{11}$, $c_0$, and $c_1$ which were equivalent to the vacuum
Einstein equations $G_{\mu\nu}=0$.  We have found a streamlined
derivation which treats the indexed quantities $\lambda_{AB}$ and
$c_A$ as single entities, rather than dealing with each component
individually.  We have also derived explicit expressions (given in
\cite{2KV}) for the components $G_{AB}$, $G_{Ai}$, and $G_{ij}$ which
can be used even when the stress-energy tensor is non-vanishing.

In the case where the off-block-diagonal components $\{T_{Ai}\}$ of
the stress-energy tensor vanish, the Einstein equations $G_{Ai}=0$
simply say that the scalars $c_A$ are constants.  A convenient choice
of gauge in which to solve the remaining six Einstein equations is a
basis in which
\begin{eqnarray}
    c_0\equiv 2\Omega
    \,
    ,
    \qquad
    c_1\equiv 0
    \,
    ,
    \\
    \lambda_{00}=(\lambda+X(\rho,\varphi)^2)Z(\rho,\varphi)^{-1}
    \,
    ,
    \quad
    \lambda_{01}=X(\rho,\varphi)
    \,
    ,
    \quad
    \lambda_{11}=Z(\rho,\varphi)
    \,
    ,
    \\
    \gamma_{22}=1
    \,
    ,
    \qquad
    \gamma_{23}=0
    \,
    ,
    \qquad
    \gamma_{33}=-\lambda(\rho,\varphi)^{-1}\rho^2 F(\rho,\varphi)
    \,.
\end{eqnarray}
These four functions $X$, $Z$, $\lambda$, and $F$ of the two {\coord}s
$x^2=\rho$ and $x^3=\varphi$ provide a fixed gauge in which the
problem can be solved numerically.  There are six block-diagonal
Einstein equations (for $\{G_{AB}\}$ and $\{G_{ij}\}$), but only four
of them are independent because of the contracted Bianchi identities
$(dx^i)_\mu \nabla_\nu G^{\mu\nu}=0$.  These equations will be
combined with a treatment of the cosmic string sources and the
boundary conditions at infinity and the origin to produce a
numerically determined spacetime.

\section{QSBI II: Radiation-Balanced Boundary Conditions \\
(Whelan, Krivan and Price \cite{radbal})}
\label{sec:radbal}

\subsection{Radiative Boundary Conditions}

The true physical spacetime which we are ultimately modelling contains
gravitational radiation at infinity which is outgoing.  This loss of
energy leads to decay of the orbits and inspiral of the compact
objects in a non-stationary solution.  In many radiative problems, a
solution with outgoing radiation can be constrained to be stationary
by an external force whose agent does not couple to the radiation.
However, this is at odds with the idea that in General Relativity all
matter gravitates, so we should look instead for a solution where
there is no net energy loss to infinity due to the radiation.  A
na\"{\i}ve replacement for the outgoing radiation boundary conditions
would be a standing wave condition.  However, that turns out to be
inappropriate, as standing waves require a node (Dirichlet) or
extremum (Neumann) at a particular location, and a standing wave
condition thus fails to converge to a well-defined limit as that
location is moved out to infinity.

We have been investigating the question of how to implement a sensible
boundary condition leading to a balance of radiation in the context of
a simple theory: a nonlinear scalar field $\psi(t,\rho,\phi)$ in
two-plus-one dimensions, where the source $\sigma$ and field $\psi$
are required to be co-rotating (i.e., depend only on $\rho$ and
$\varphi=\phi-\Omega t$).  This causes the field equation to take the
form
\begin{equation}
  \label{coroteqn}
  \Box^2\psi=
  \frac{\partial^2\psi}{\partial\rho^2}
  +\rho^{-1}\frac{\partial\psi}{\partial\rho}
  +\left(
    \rho^{-2}-\Omega^2 c^{-2}
  \right)    
  \frac{\partial^2\psi}{\partial\varphi^2}
  = \sigma + \lambda \psi^3 
  .
\end{equation}

In the numerical solution for the non-linear equation
\eqref{coroteqn}, one uses a finite {\coord} grid and specifies a set
of boundary conditions at the large finite radius $\rho=R$ which marks
the end of the grid.  Table \ref{tab:1}
\begin{table}[tbp]
\label{tab:1}
  \begin{tabular}{r|c|c}
    &BCs at $\rho=R$ & $R\longrightarrow\infty$ limit  \\
    \hline
    Outgoing & $(\partial_\rho+c^{-1}\partial_t)\psi^{\text{out}}_R=0$ &
 $\psi^{\text{out}}\sim e^{im(\Omega
    c^{-1}\rho+\varphi)}\sim e^{im\Omega(c^{-1}\rho-t)}$ \\
 Ingoing &
  $(\partial_\rho-c^{-1}\partial_t)\psi^{\text{in}}_R=0$ &
  $\psi^{\text{in}}\sim e^{im(\Omega
    c^{-1}\rho-\varphi)}\sim e^{im\Omega(c^{-1}\rho+t)}$ \\
 Neumann SW &
  $\partial_\rho\psi^{\text{N}}_R=0$ &
 {\bf N/A} \\
 Dirichlet SW &
  $\psi^{\text{D}}_R=0$ &
 {\bf N/A} \\
  \end{tabular}
\caption{Some large-distance boundary conditions.  
  Purely ingoing or outgoing radiation has a well-defined limit as the
  radius $R$ at which it is applied is taken to infinity, but standing
  wave conditions (e.g., Neumann or Dirichlet) do not.}
\end{table}
shows several possible boundary conditions.  We call, e.g., the
solution resulting from the application of the outgoing boundary
condition at a particular $R$, $\psi^{\text{out}}_R$.  The local
conditions defining outgoing or ingoing radiation each produce
well-defined limits (which we call simply $\psi^{\text{out}}$ and
$\psi^{\text{in}}$) as $R$ is taken to infinity, but the standing wave
solutions do not converge to any limit.

Taking the average of the $R\rightarrow\infty$ limits
$\psi^{\text{out}}$ and $\psi^{\text{in}}$ defines a function
\begin{equation}
  \psi^{\text{avg}}=\frac{1}{2}(\psi^{\text{in}}+\psi^{\text{out}})
\end{equation}
which has a balance of ingoing and outgoing radiation without
reference to any particular radius $R$.  In the linear ($\lambda=0$)
theory, where the principle of superposition holds, it is a solution
to the field equations, but in the non-linear theory it is not.  It is
also not defined as the limit of any local boundary conditions (in
particular, the is average of out- and ingoing solutions is not
produced by an average of out- and ingoing boundary conditions).

\subsection{Green's Function Solution}

The tasks of defining a solution analogous to $\psi^{\text{avg}}$ and
numerically determining that solution in the absence of a local
boundary condition can both be accomplished by a Green's function
method.

If we define a Green's function $G(\rho,\varphi|\rho_0,\varphi_0)$
such that
\begin{equation}
  \label{GFeqn}
  \Box^2 G = \rho_0^{-1}\delta(\rho-\rho_0)\,\delta(\varphi-\varphi_0)
  \,
  ,
\end{equation}
the differential equation \eqref{coroteqn} can be converted into an
integral equation
\begin{equation}
  \label{NLinteqn}
  \psi(\rho,\varphi)=\int\rho_0\,d\rho_0\,d\varphi_0\,
  G(\rho,\varphi|\rho_0,\varphi_0) \,
  [\sigma(\rho_0,\varphi_0)+\lambda\psi^3(\rho,\varphi)]
  \,
  .
\end{equation}
Solving to the Green's function equation \eqref{GFeqn} requires the
application of boundary conditions at infinity, so there are actually
a family of Green's functions with boundary conditions ``built into''
them.  For example, using the retarded Green's function
$G^{\text{out}}$ in \eqref{NLinteqn} gives the outgoing solution
$\psi^{\text{out}}$, using the advanced Green's function
$G^{\text{in}}$ gives the ingoing solution $\psi^{\text{in}}$, and
similarly for any of the boundary conditions defined at finite radii
in Table~\ref{tab:1}.

Since the Green's function equation \eqref{GFeqn} is linear, even when
the wave equation \eqref{coroteqn} is not, we can always average the
retarded and advanced \emph{Green's functions} to give a
time-symmetric Green's function
\begin{equation}
  G^{\text{sym}}=\frac{1}{2}(G^{\text{in}}+G^{\text{out}})
\end{equation}
which defines a radiation-balanced solution $\psi^{\text{sym}}$.

In the linear theory, this solution $\psi^{\text{sym}}$ will be equal
to the average $\psi^{\text{avg}}$ of the ingoing and outgoing
solutions.  In the non-linear theory, we can calculate the two
numerically (by iterating \eqref{NLinteqn}) and compare them.  We are
ultimately trying to use the outgoing piece of $\psi^{\text{sym}}$ as
an approximation for $\psi^{\text{out}}$ (and thus as a further
approximation for the actual inspiralling solution), and we expect that
approximation to be good when
$\psi^{\text{sym}}\approx\psi^{\text{avg}}$.

\subsection{Numerical Results}

\begin{figure}[tbp]
    \epsfig{file=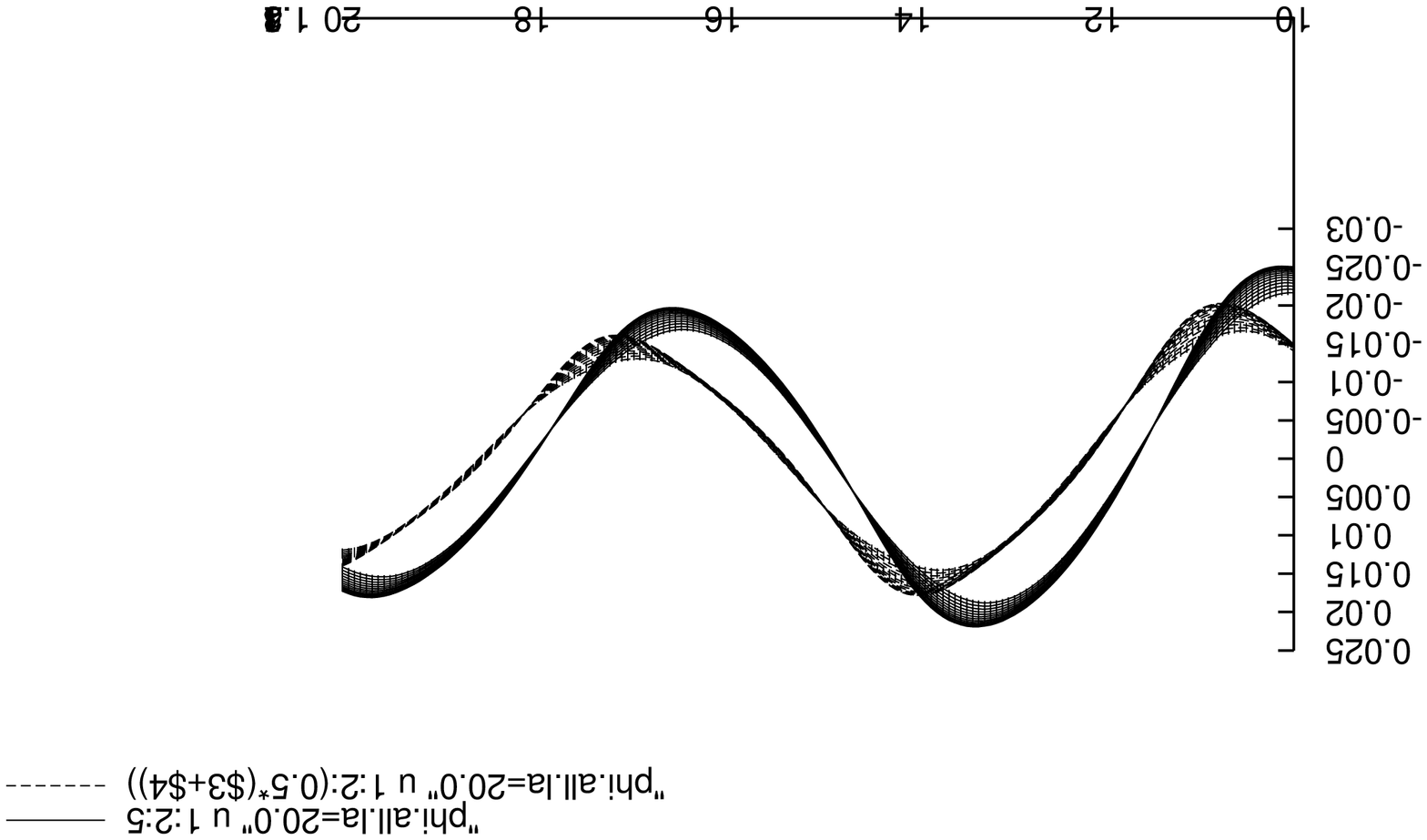,width=15pc,clip=,angle=90,%
bbllx=430,bblly=160,bburx=160,bbury=610}      
  \begin{center}
    \caption{Comparison between 
      $\psi^{\text{sym}}$ and $\psi^{\text{avg}}$ in the wave zone for
      a source of two particles of opposite charge at $\rho/2=.25$,
      $\varphi=(\pi\pm\pi)/2$, with a rotational frequency of
      $\Omega=1/2$ and an orbital velocity of one-fourth the speed of
      light.  The fields are plotted versus $\rho/2$ for the range of
      angles $\varphi\in[1,\pi/2]$.  The amplitude of
      $\psi^{\text{avg}}$ is around 20\% smaller and the phase is
      shifted 40 degrees relative to $\psi^{\text{sym}}$.  This
      agreement is only middling because the sources are still
      somewhat relativistic.}
    \label{fig:.25c}
  \end{center}
\end{figure}

\begin{figure}[tbp]
    \epsfig{file=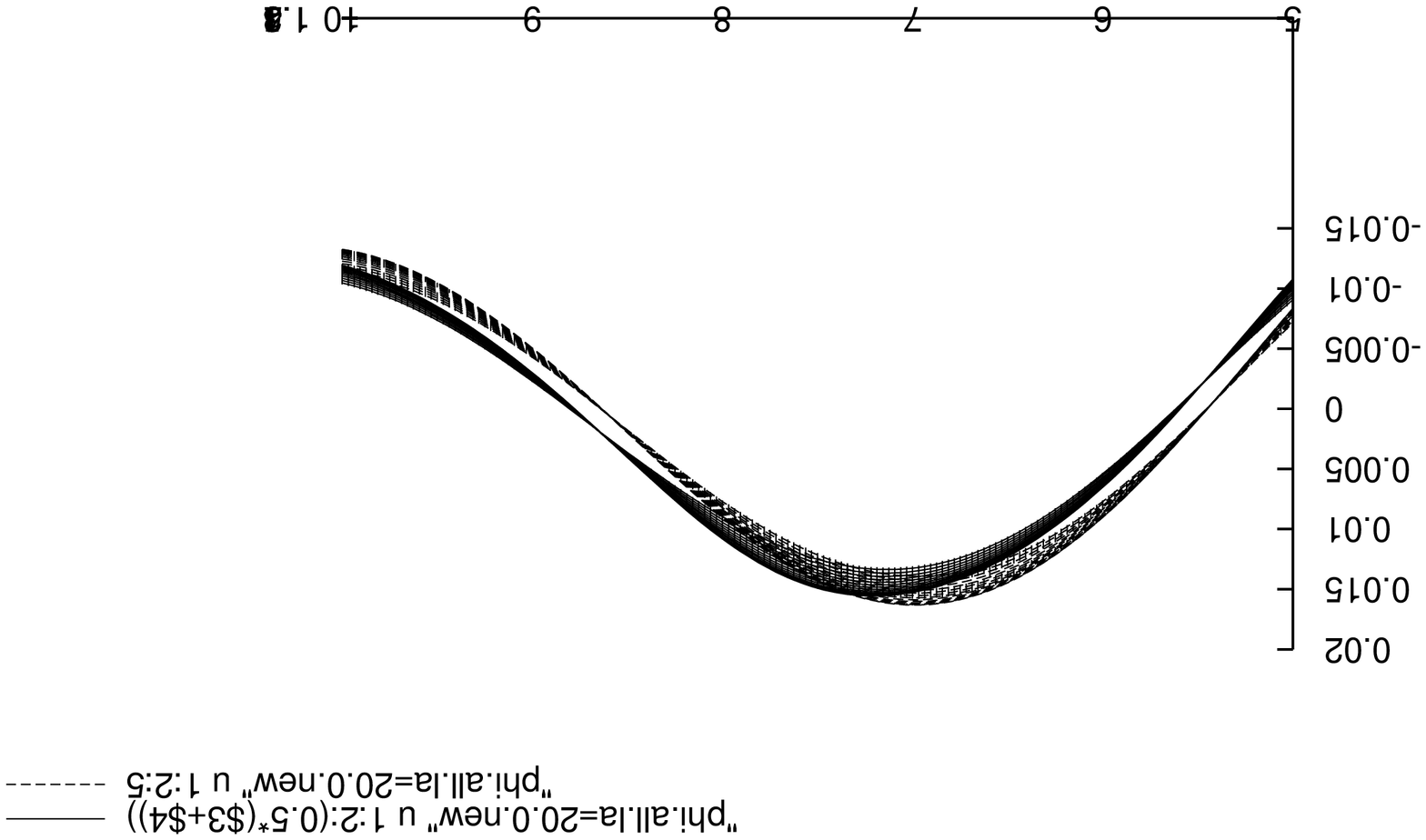,width=15pc,clip=,angle=90,%
bbllx=430,bblly=160,bburx=160,bbury=610}      
  \begin{center}
    \caption{Comparison between 
      $\psi^{\text{sym}}$ and $\psi^{\text{avg}}$ in the wave zone for
      a source of two particles of opposite charge at $\rho/2=.125$,
      $\varphi=(\pi\pm\pi)/2$, with a rotational frequency of
      $\Omega=1/2$ and an orbital velocity of one-eighth the speed of
      light.  The fields are plotted versus $\rho/2$ for the range of
      angles $\varphi\in[1,\pi/2]$.  With these less relativistic
      sources, the agreement is better than in Figure~\ref{fig:.25c},
      with approximately a 9\% smaller amplitude for
      $\psi^{\text{avg}}$ and 7 degree phase shift.}
    \label{fig:.125c}
  \end{center}
\end{figure}

Numerical calculations \cite{radbal} for a strongly non-linear
($\lambda=20$) theory have shown that the approximation becomes good
as the orbital velocity of the sources becomes non-relativistic.
Figures \ref{fig:.25c} and \ref{fig:.125c}
show the comparison between
$\psi^{\text{sym}}$ and $\psi^{\text{avg}}$ for orbital velocities of
one-fourth and one-eighth the speed of light, respectively.

\begin{figure}[tbp]
    \epsfig{file=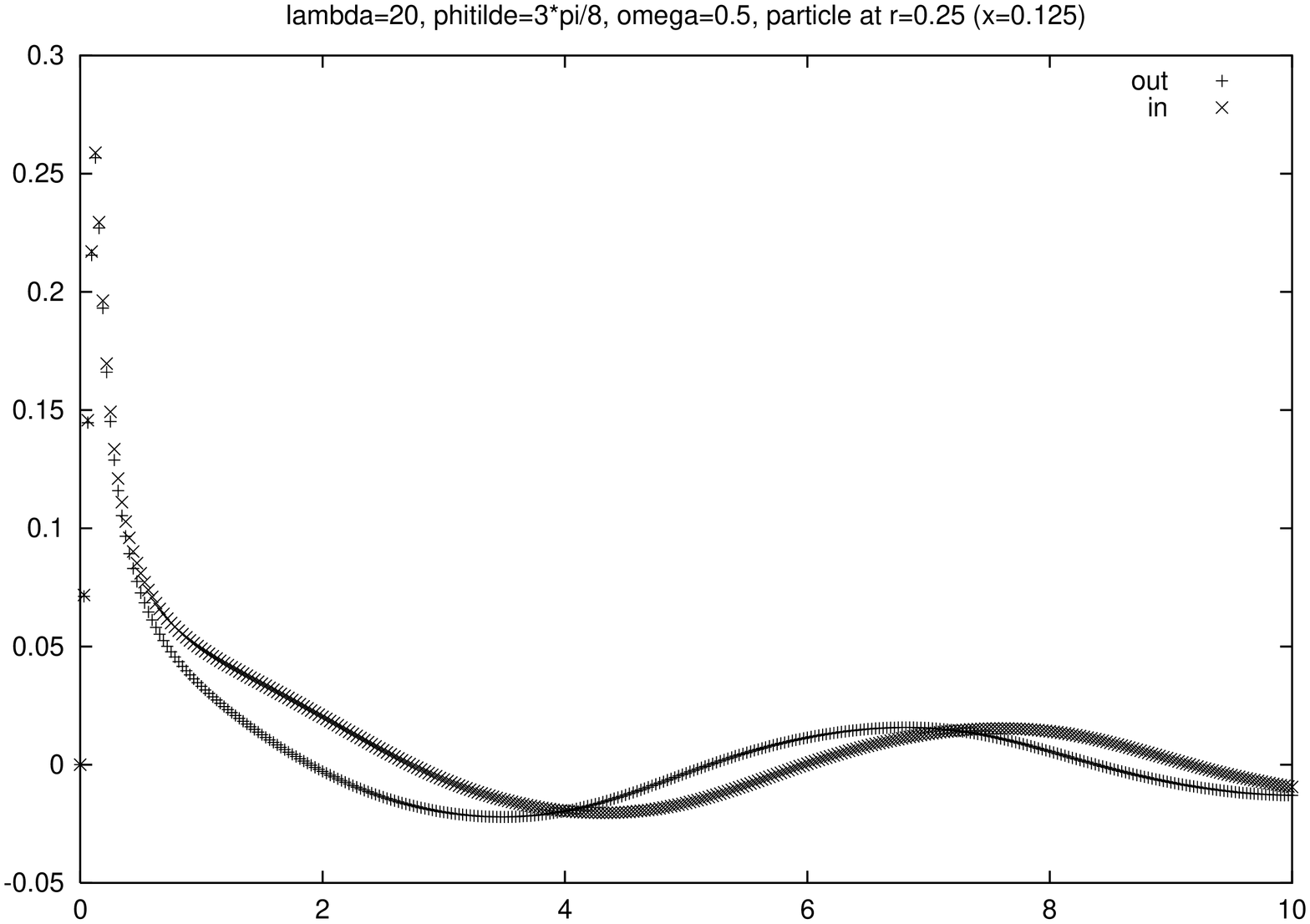,width=20pc,clip=,angle=-90,%
bbllx=75,bblly=90,bburx=540,bbury=770}
  \begin{center}
\caption{Illustration of the agreement between $\psi^{\text{out}}$ and
  $\psi^{\text{in}}$ in the strong-field region.  The parameters are
  the same as in Figure \ref{fig:.125c}, i.e., oppositely-charged
  source particles at $\rho/2=.125$, $\varphi=(\pi\pm\pi)/2$, a
  rotational frequency of $\Omega=1/2$ and an orbital velocity of
  one-eighth the speed of light.  The fields are plotted against
  $\rho/2$ for $\varphi=3\pi/8$.  Note that while the two solutions
  differ in the outer, wave zone, they are nearly equal in the inner
  regions which are the source of the most of the nonlinear effects,
  leading to the approximate agreement of $\psi^{\text{sym}}$ and
  $\psi^{\text{avg}}$ (which are identical in the linear theory)
  illustrated in Figure \ref{fig:.125c}.}
\label{fig:try}
\end{center}
\end{figure}

This agreement occurs even when the theory is highly nonlinear
because, even though $\psi^{\text{in}}$ and $\psi^{\text{out}}$ differ
in the wave zone, they are approximately the same in the inner,
strong-field region, as illustrated in Figure \ref{fig:try}.

The final step in the scalar field theory analysis will be to
numerically extract the ``outgoing part'' from the time-symmetric
solution $\psi^{\text{sym}}$ and compare it directly to
$\psi^{\text{out}}$; we expect that the two will agree when
$\psi^{\text{sym}}$ is approximately equal to $\psi^{\text{avg}}$.
  
\section{QSBI III: Gravitational Waves on a Cosmic String Background} 
\label{sec:cyl}

\subsection{Schwarzschild vs.\ Levi-Civita Backgrounds}

In the three-plus-one binary inspiral problem, with non-extended
sources, gravitational waves at infinity can be treated as
perturbations to Minkowski spacetime.  This is because at large distances,
where higher multipoles of the non-radiative field can be ignored,
the non-radiative part of the geometry is associated with the Schwarzschild
metric
\begin{equation}
-(1-r_s/r)\,c^2dt^2+(1-r_s/r)^{-1}dr^2+r^2d\Omega^2
\,
,
\end{equation}
which contains a length scale $r_s=2GM/c^2$; taking
$r/r_s\rightarrow\infty$ reduces the Schwarzschild line element to the
Minkowski one, which is just another way of saying that the
Schwarzschild solution is asymptotically flat.

With infinitely extended sources, this is not possible.  Far from the
strings, where the internal structure is not felt, the non-radiative
effects are approximated by the Levi-Civita solution \cite{cs} of a
single cosmic string at the origin with mass-per-unit length $\Lambda$, whose
line element is given for small $\Lambda$ by
\begin{equation}
  - \rho^{2C} c^2 dt^2 + \rho^{-2C} dz^2 + d\rho^2 + \rho^2 d\phi^2
  \,
  .
\end{equation}
The mass parameter $C=2G\Lambda/c^2$ is now dimensionless; this lack
of a length scale means that Levi-Civita spacetime is \emph{not}
asymptotically flat and we must use Levi-Civita rather than Minkowski
as our background metric.

\subsection{Perturbations to Levi-Civita}

Suppose the metric is given by the Levi-Civita metric plus a small
perturbation $h_{\mu\nu}(t,\rho,\phi)$.  (In our problem the $t$ and
$\phi$ dependence will further be restricted to a function of
$\phi-\Omega t$, but we are thinking here of more general
considerations.)  We can divide the components $\{h_{\mu\nu}\}$ into
those which are odd under inversion of the $z$ {\coord} ($h_{zt}$,
$h_{z\rho}$, and $h_{z\phi}$) and those which are even ($h_{zz}$,
$h_{tt}$, $h_{t\rho}$, $h_{t\phi}$, $h_{\rho\rho}$, $h_{\rho\phi}$,
and $h_{\phi\phi}$).  In the linearized theory, the odd and even
sectors are uncoupled.

To count the physical degrees of freedom, we consider the effects of a
gauge transformation
\begin{equation}
  h_{\mu\nu}\rightarrow h_{\mu\nu}-2\,\xi_{(\mu;\nu)}
\end{equation}
by a gauge parameter $\xi_\mu(t,\rho,\phi)$, as well as the
Hamiltonian constraint $H$ and the three components
$\{H_z,H_\rho,H_\phi\}$ of the momentum constraint.

The three odd components of the metric perturbation consist of one
gauge degree of freedom $\xi_{z}$, one constraint $H_{z}$, and one
physical degree of freedom, which can be thought of as the ``cross
polarization'' (since it involves $h_{z\phi}$).

The seven even components include three gauge degrees of freedom
$\xi_{t}$, $\xi_{\rho}$, and $\xi_{\phi}$, three constraints $H$,
$H_{\rho}$, and $H_{\phi}$, and one physical ``plus polarization''
involving $h_{zz}$ and $h_{\phi\phi}$.

The next step in a treatment akin to that done by Moncrief
\cite{moncrief} for waves on a Schwarzschild background will be to
derive explicit wave equations for these two gauge-invariant physical
degrees of freedom.  For the purposes of the quasi-stationary binary
inspiral program, it may suffice to work with linearized versions of
the gauge-fixed equations derived in \cite{2KV}, but the picture of
general perturbations to Levi-Civita will at least be useful in
interpreting the quantities involved.

\section{Future Outlook}

The work described in Section~\ref{sec:2KV} has produced non-linear
differential equations which can be numerically implemented given a
suitable treatment of the sources and the boundary conditions on the
radiation at the outer boundary.  The study of radiation balance in
the context of non-linear scalar field theory in
Section~\ref{sec:radbal} will be completed by the extraction of the
outgoing part of the time-symmetric Green's function solution and the
comparison between that and the actual outgoing solution (which we
expect to be unavailable in the gravitational case).  Application of
these methods to General Relativity will be complicated by the fact
that it is probably not possible to formulate a (linear) Green's
Function solution to the non-linear wave equation in that case.  The
analysis of perturbations to Levi-Civita spacetime described in
Section~\ref{sec:cyl} is still in the early stages, and its full
development may or may not be necessary to the understanding of our
problem, which possesses the additional co-rotational Killing vector.

Finally, should the technique of finding a stationary solution with a
balance of ingoing and outgoing radiation prove successful in the
essentially two-dimensional problem of orbiting cosmic strings, it can
be applied to the computationally more intensive three-dimensional
problem.

\section*{Acknowledgments}

The author would like to thank his collaborators J.~D.~Romano,
W.~Krivan, R.~H.~Price, as well as C.~Torre, K.~Thorne, J.~Creighton,
J.~Friedmann, S.~Morsink, A.~Held, and the Relativity Group at the
University of Utah, where this work was begun, and acknowledge the
financial support of NSF Grant PHY-9734871, Swiss Nationalfonds, and
the Tomalla Foundation Z\"{u}rich.

\section*{References}

%%% Standard "thebibliography" LaTeX environment. References sorted BY
%%% ORDER OF APPEARANCE for hypertext links and cross-referencing.
%%%

\end{document}